# Exact Solutions and Excitations for the Davey-Stewartson Equations with Nonlinear and Gain Terms

Ren-Jie Wang[1,a] and Yong-Chang Huang[1, 2, 3,b]

1. Institute of Theoretical Physics, Beijing University of Technology, Beijing, 100124, China
2. Kavli Institute for Theoretical Physics, Chinese Academy of Sciences, Beijing, 100190, China
3. CCAST (World Lab.), P. O. Box 8730, Beijing, 100080, China

**Abstract**

We study the general (2+1)-dimensional Davey-Stewartson (DS) equations with nonlinear and gain terms and acquire explicit solutions through variable separation approach. In particular, we deduce some main novel excitations for the DS equations, and further demonstrate different features of these excitations. More importantly, the similar solutions and excitations can be predicted to exist in other related revolution equations such as nonlinear Schrödinger equation to explain the Bose-Einstein condensation.

## I. INTRODUCTION

It is well known that nonlinear equations play a crucial role in the study of nonlinear physical phenomena, such as plasma [1], fluid dynamics, nonlinear optics, elastic media, etc. Solving these nonlinear equations may lead us to understand the described procedures profoundly and guide us to identify the facts that are not simply comprehended through common observations. In recent years, many experts have developed some explicit methods, for example, the inverse scattering method [2, 3], Bäcklund transformation [4, 5], Darboux transformation [4], Cole–Hopf transformation, tanh-coth method [6], Hirota bilinear method [6-8], Painlevé expansion method [9] and so on.

The Davey-Stewartson (DS) system [10] is a system for the evolution of weakly nonlinear packets of water waves in the finite depth that travel in one direction but in which the amplitude of waves is modulated in two spatial directions. In a lot of two-dimensional systems, both short waves and long waves may coexist, and an accurate depiction of two-dimensional modulation of nonlinear waves ought to involve both short wave and long wave modes. The DS system is just such a system. Furthermore, DS equations, a pair of coupled nonlinear equations in two dependent variables, can be reduced to the (1+1)-dimensional nonlinear Schrödinger (NLS) equation by carrying out a suitable dimensional reduction.

Several analytical solutions have considerably been developed to be used for nonlinear partial differential equations such as DS equations that have special kinds of solutions in recent years, for instance, the growing-and-decaying mode solutions [11-13], dromions, breathers, instantons, propagating and periodic wave patterns [14, 15].

Our paper is organized as follows: in Sect. II, we consider the (2+1)-dimensional DS equations and obtain exact and explicit solutions through variable separation approach; in Sect. III, we deduce and demonstrate different excitations for DS equations; Sect. IV gives out the summary and results of this paper. Furthermore, we utilize the achieved conclusions to address more on how the excitation patterns predicted in this paper can be observed in some interesting systems, e. g. Bose-Einstein condensation and nonlinear optical systems.

---

[a] Corresponding Author, E-mail: roger_wang@emails.bjut.edu.cn, r.jiewang@gmail.com
[b] E-mail: ychuang@bjut.edu.cn

## II. EXACT SOLUTIONS FOR THE DAVEY-STEWARTSON EQUATIONS

We generalize the constant-coefficient Davey-Stewartson (DS) equation in (2+1)-dimensions [16] to (2+1)-dimensional variable-coefficient Davey-Stewartson (DS) equations

$$\begin{cases} i\partial_t u + \beta(t)[\frac{1}{2}(u_{xx}+u_{yy}) - \frac{1}{2}|u|^2 u + u\phi] = i\dot{\gamma}(t)u, \\ \phi_{xx} - \phi_{yy} = (uu^*)_{xx}, \end{cases} \quad (1)$$

where function $\beta(t)$ stands for dispersion and nonlinearity coefficients, $\gamma(t)$ represents gain coefficients, $u$ is the complex wave envelope and the real function $\phi$ might be regarded as a forcing term of the first equation of equation group (1). When $\gamma(t)=0$ and $\beta(t)=const.$, equation group (1) becomes the ordinary DS equations [16]. In absence of $y$ dependence, the DS equations reduce to the (1+1)-dimensional nonlinear Schrödinger (NLS) equation, i.e.

$$i\partial_t u + \frac{\beta(t)}{2} u_{xx} + \frac{1}{2}\beta(t)|u|^2 u = i\dot{\gamma}(t)u. \quad (2)$$

Now we apply variable separation approach [17] for the (2+1)-dimensional DS equations (1), which can be rewritten as

$$\begin{cases} i\partial_t u + \beta(t)[\frac{1}{2}(u_{\zeta\zeta}+u_{\eta\eta}) - \frac{1}{2}|u|^2 u + u\phi] = i\dot{\gamma}(t)u, \\ 4\phi_{\zeta\eta} = (uu^*)_{\zeta\zeta} + (uu^*)_{\eta\eta} + 2(uu^*)_{\zeta\eta}, \end{cases} \quad (3)$$

with coordination transformations $\zeta = \frac{x-y}{\sqrt{2}}, \eta = \frac{x+y}{\sqrt{2}}$. To deduce exact solutions of the DS equations, we can introduce dependent variable transformations

$$\begin{cases} u = \dfrac{g}{f}, \\ \phi = 2(\log f)_{xx} + p_0(\zeta,t) + q_0(\eta,t) \\ \quad = -\dfrac{(f_\zeta+f_\eta)^2}{f^2} + \dfrac{f_{\zeta\zeta}+2f_{\zeta\eta}+f_{\eta\eta}}{f} + p_0(\zeta,t) + q_0(\eta,t), \end{cases} \quad (4)$$

where functions $f$ and $g$ are real and complex, respectively, and $p_0 = p_0(\zeta,t), q_0 = q_0(\eta,t)$ are arbitrary functions.

Therefore, the bilinear equations for equation group (3) are

$$\begin{cases} ifD_t g \cdot f + \dfrac{\beta}{2} f(D_\zeta^2 + D_\eta^2)g \cdot f - \dfrac{\beta}{2} g^2 g^* + \beta g(D_\zeta D_\eta + p_0 + q_0)f \cdot f - i\dot{\gamma} g \cdot f^2 = 0, \\ 2D_\zeta D_\eta f \cdot f - gg^* = 0, \end{cases} \quad (5)$$

in which the well-known Hirota bilinear operators are defined by the rule [18]

$$D_x^m D_t^n a \cdot b = (\frac{\partial}{\partial x} - \frac{\partial}{\partial x'})^m (\frac{\partial}{\partial t} - \frac{\partial}{\partial t'})^n a(x,t) \cdot b(x',t')\big|_{x'=x, t'=t}. \quad (6)$$

On the other hand, because a lot of general physical processes should satisfy quantitative causal relation with no-loss-no-gain character [19-21], e.g., Ref.[22] uses the no-loss-no-gain homeomorphic map transformation satisfying the quantitative causal relation to gain exact strain tensor formulas in Weitzenbock manifold. In fact, some changes ( cause ) of some quantities in the first equation of (5) must result in the relative some changes ( result ) of the other quantities in the first equation of (5) so that the right side of the first equation of (5) keeps no-loss-no-gain, i.e., zero, namely, the first equation of (5) also satisfies the quantitative causal relation, so does the second equation of (5).

We now utilize the variable separation approach to solve Eq. (5), which is applicable to a large of nonlinear equations. Substituting following expansions

$$f = a_0 + a_1 p(\zeta,t) + a_2 q(\eta,t) + a_3 p(\zeta,t)q(\eta,t) \text{ and } g = p_1(\zeta,t)q_1(\eta,t)\exp\{ir(\zeta,t)+is(\eta,t)\} \quad (7)$$

into equation group (5), we can obtain a series of equations

$$\begin{cases} (a_0 a_3 - a_1 a_2) p_\zeta q_\eta = \dfrac{1}{4} p_1^2 q_1^2, \\ (a_0 + a_1 p + a_2 q + a_3 pq)[p_{1t}q_1 + p_1 q_{1t} + \dfrac{\beta}{2}(2p_{1\zeta}q_1 r_\zeta + p_1 q_1 r_{\zeta\zeta} + 2p_1 q_{1\eta} s_\eta + p_1 q_1 s_{\zeta\zeta}) - \gamma p_1 q_1] \\ \qquad = p_1 q_1 (a_1 + a_3 q)(p_t + \beta r_\zeta p_\zeta) + p_1 q_1 (a_2 + a_3 p)(q_t + \beta s_\eta q_\eta), \\ -p_1 q_1 (r_t + s_t) + \dfrac{\beta}{2}[(p_{1\zeta\zeta}q_1 - p_1 q_1 r_\zeta^2) + (p_1 q_{1\eta\eta} - p_1 q_1 s_\zeta^2)] + \beta p_1 q_1 (p_0 + q_0) = 0, \end{cases} \quad (8)$$

where functions $p = p(\zeta,t), q = q(\eta,t), p_1 = p_1(\zeta,t), q_1 = q_1(\eta,t), r = r(\zeta,t), s = s(\eta,t)$ are all real, $a_0, a_1, a_2, a_3$ are arbitrary constants. According to the fact that $\{p, p_0, p_1, r\}$ and $\{q, q_0, q_1, s\}$ are only functions of $\{\zeta, t\}$ and $\{\eta, t\}$, respectively, we can separate different variables into corresponding different sides of each equation in equation group (8), as a result, we achieve the following variable separated equations

$$\begin{cases} p_1 = \delta_1 \sqrt{c_0 p_\zeta}, \\ q_1 = 2\delta_2 \sqrt{c_0^{-1}(a_0 a_3 - a_1 a_2)q_\eta}, \quad (\delta_1^2 = \delta_2^2 = 1) \end{cases} \quad (9)$$

$$\begin{cases} p_t = -\beta r_\zeta p_\zeta + c_1 (a_2 + a_3 p)^2 + c_2 (a_2 + a_3 p) - (a_0 a_3 - a_1 a_2) c_3, \\ q_t = -\beta s_\eta q_\eta - c_3 (a_1 + a_3 q)^2 - c_2 (a_1 + a_3 q) + (a_0 a_3 - a_1 a_2) c_1, \end{cases} \quad (10)$$

$$\begin{cases} p_0 = \dfrac{1}{\beta}[c_5 + r_t - \dfrac{\beta}{2}(\dfrac{p_{1\zeta\zeta}}{p_1} - r_\zeta^2)], \\ q_0 = \dfrac{1}{\beta}[-c_5 + s_t - \dfrac{\beta}{2}(\dfrac{q_{1\eta\eta}}{q_1} - s_\zeta^2)], \end{cases} \quad (11)$$

where $c_i = c_i(t)$ $(i = 0,3,4,5)$ are arbitrary functions, while functions $c_1(t)$ and $c_2(t)$ satisfy

$$\begin{cases} c_1 = \dfrac{1}{a_3(a_2 + a_3 p)}[\dfrac{p_{1t}}{p_1} + \beta(\dfrac{p_{1\zeta}r_\zeta}{p_1} + \dfrac{r_{\zeta\zeta}}{2}) - \gamma - c_4], \\ c_2 = \dfrac{-1}{a_3(a_1 + a_3 q)}[\dfrac{q_{1t}}{q_1} + \beta(\dfrac{q_{1\eta}s_\eta}{q_1} + \dfrac{s_{\eta\eta}}{2}) + c_4]. \end{cases} \quad (12)$$

Because the arbitrary functions $p_0$ and $q_0$ can be represented by $p$ and $q$ in equation group (11), we can treat both $p$ and $q$ as arbitrary functions of $\{\zeta, t\}$ and $\{\eta, t\}$, respectively. Consequently, we obtain the solution of DS equation group (1)

$$u = \frac{2\delta_1 \delta_2 \sqrt{(a_0 a_3 - a_1 a_2) p_\zeta q_\eta} \exp\{ir + is\}}{a_0 + a_1 p + a_2 q + a_3 pq}. \tag{13}$$

The intensity of the wave envelope is given by the field

$$U = |u|^2 = \frac{4(a_0 a_3 - a_1 a_2) p_\zeta q_\eta}{(a_0 + a_1 p + a_2 q + a_3 pq)^2}. \tag{14}$$

According to (14), we recognize that the excitations are also dependent on the selections of $\{p, q\}$, since $p(\zeta, t), q(\eta, t)$ are arbitrary functions, the excitations of the field $U$ are so various.

On the other hand, we need to avoid the singularities of selections, which equates $a_0 + a_1 p + a_2 q + a_3 pq = 0$, and the selection of $\{p, q\}$ should also satisfy the relation $(a_0 a_3 - a_1 a_2) p_\zeta q_\eta > 0$ due to (14) > 0.

## III. EXCITATIONS FOR THE DAVEY-STEWARTSON EQUATIONS

According to the general solution (14), the excitations of field $U$ have different patterns, which are dependent on the selections of $\{p, q\}$. Now we discuss these excitation and their characters.

**Case 1:** Dromion and solitoff excitations

Here we generally select $\{p, q\}$ as

$$\begin{cases} p(\zeta, t) = \sum_{i=1}^{N} A_i \cdot \exp\{K_i \zeta + L_i t + \vartheta_i^0\}, \\ q(\eta, t) = \sum_{j=1}^{M} B_j \cdot \exp\{G_j \eta + H_j t + \theta_j^0\}, \end{cases} \tag{15}$$

where $A_i, B_j, K_i, G_j, L_i, H_j, \vartheta_i^0, \theta_j^0$ are arbitrary constants and $M, N$ are arbitrary positive integers, thus we can obtain some main excitation patterns, such as dromion patterns, solitoff patterns, resonant solitoff patterns and so on.

When choosing $p = \exp\{\zeta + t + 1\}$, $q = \exp\{\eta + t + 1\}$, the intensity of the wave envelope, i.e. the field $U$, indicates a single dromion excitation, which is showed by Fig. 1. (a).

When picking $p = \exp\{\zeta + t + 1\} + \exp\{2\zeta + t + 1\}$, $q = \exp\{\eta - t + 1\} + \exp\{2\eta - t + 1\}$, the excitation for the field $U$ is a single solitoff pattern, Fig. 1. (b) portrays this excitation which looks like a single-line-soliton.

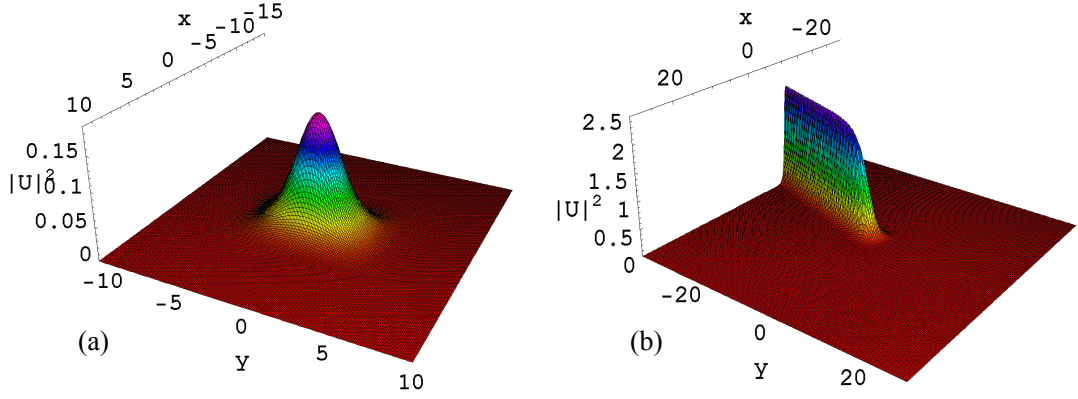

Fig. 1. (Color online) (a) is a single dromion excitation with parameters $K_i = G_j = L_i = H_j = \vartheta_i^0 = \theta_j^0 = A_i = B_j = M = N = 1$, $(i = j = 1)$ and $a_0 = 1, a_1 = 1, a_2 = 1, a_3 = 2$; (b) is a single solitoff pattern excitation, its parameters are $K_1 = G_1 = L_1 = L_2 = \vartheta_1^0 = \vartheta_2^0 = \theta_1^0 = \theta_2^0 = A_1 = A_2 = B_1 = B_2 = 1$, $K_2 = G_2 = 2, H_1 = H_2 = -1$. and $a_0 = 2, a_1 = 0, a_2 = 2, a_3 = 2$. Both of these two pictures have time $t=0$.

When we select $p = -\exp\{-\zeta + t + 1\} + \exp\{2\zeta + t + 1\}$, $q = -\exp\{-\eta - t + 1\} + \exp\{2\eta - t + 1\}$, the excitation for the field $U$ is so different, Fig. 2 portrays a resonant solitoff excitation, whose shape changes along with the time. From Fig.2, when changing along with time one can find that the resonant solitoff excitations have different directions.

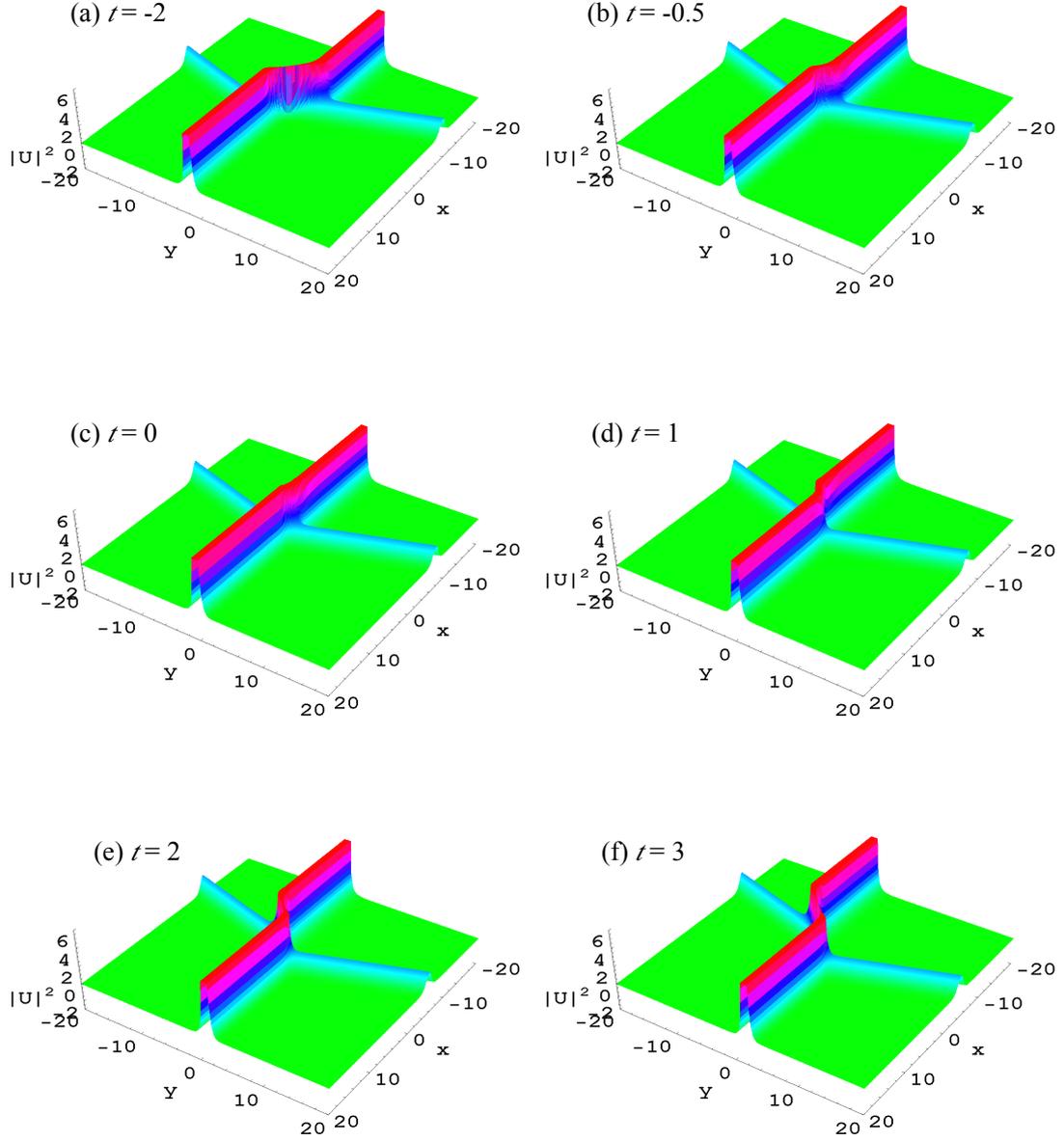

Fig. 2. (Color online) The resonant solitoff excitations. The parameters are $a_0 = 2, a_1 = -1, a_2 = 2, a_3 = 0$ and $A_1 = B_1 = K_1 = G_1 = H_1 = H_2 = -1, K_2 = G_2 = 2, A_2 = B_2 = L_1 = L_2 = \vartheta_1^0 = \vartheta_2^0 = \theta_1^0 = \theta_2^0 = 1$.

**Case 2:** Breather excitation

The diverse selections $\{p, q\}$ determine the differences of excitations, if we change the selection as

$$\begin{cases} p(\zeta, t) = 1 + \exp\{\zeta \cdot \cos^2(t)\}, \\ q(\eta, t) = \exp\{\eta + \cos^2(t)\}, \end{cases} \tag{16}$$

we will have a bizarre excitation called breather. A breather is a solitonic structure, which is also a localized periodic solution of either continuous media equations or discrete lattice equations. There are two types of breathers: standing or traveling ones. Standing breathers correspond to localized solutions whose amplitudes vary in time. The exactly solvable sine-Gordon equation [23, 24] and the focusing nonlinear Schrödinger equation [25, 26] are examples of one-dimensional partial differential equations that possess breather solutions.

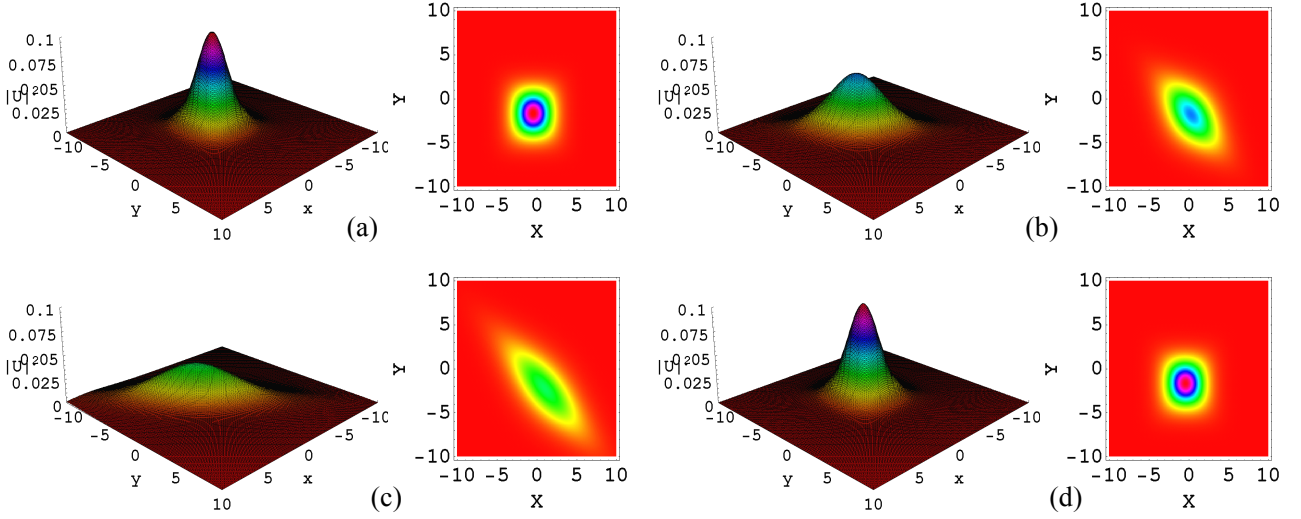

Fig. 3. (Color online) Breathes. The parameters for the field of (12) are $a_0 = 2, a_1 = 1, a_2 = 2, a_3 = 2$ and (a), (b), (c), (d) are at time $t=0, t=0.7, t=0.9, t=3$, respectively.

In Fig. 3 we can observe that the shape of field $U$ changes along with the time $t$ in one period. The right sides are the figures of projections, which are so clearly to see the variances of shapes from a circle to an ellipse and from an ellipse to a circle. The period of this breather excitation is $\pi$.

**Case 3:** Periodic wave pattern excitation

In recent years, the excitations of periodic wave patterns have been studied by many experts, for instance, the propagating, doubly periodic wave patterns [14, 15], and the long wave limits of the above special solutions and so on. In this paper, we find a new and simple periodic wave pattern, which is different from those above.

If we vary the choice of $\{p, q\}$, we have

$$\begin{cases} p(\zeta,t) = 1 + \exp\{\tan(\zeta) \cdot \cos(t) + 1\}, \\ q(\eta,t) = 1 + \exp\{\tan(\eta) \cdot \cos(t) + 1\}. \end{cases} \quad (17)$$

In Fig. 4, we can realize that the field $U$ changes along with the time periodically, when $t=0$, the field $U$ has the minimum and soon will boost at time $t=\pi/4$ and $t=\pi/3$. When it is beyond the time $t=\pi/2$, there are several tiny differences between (c) and (d), (b) and (e), (a) and (f), which have time $t=\pi/3$, $t=2\pi/3$, $t=\pi/4$, $t=3\pi/4$, $t=0$ and $t=\pi$, respectively. The field $U$ diminishes until time $t=\pi$. The period of this periodic wave pattern excitation is $\pi$.

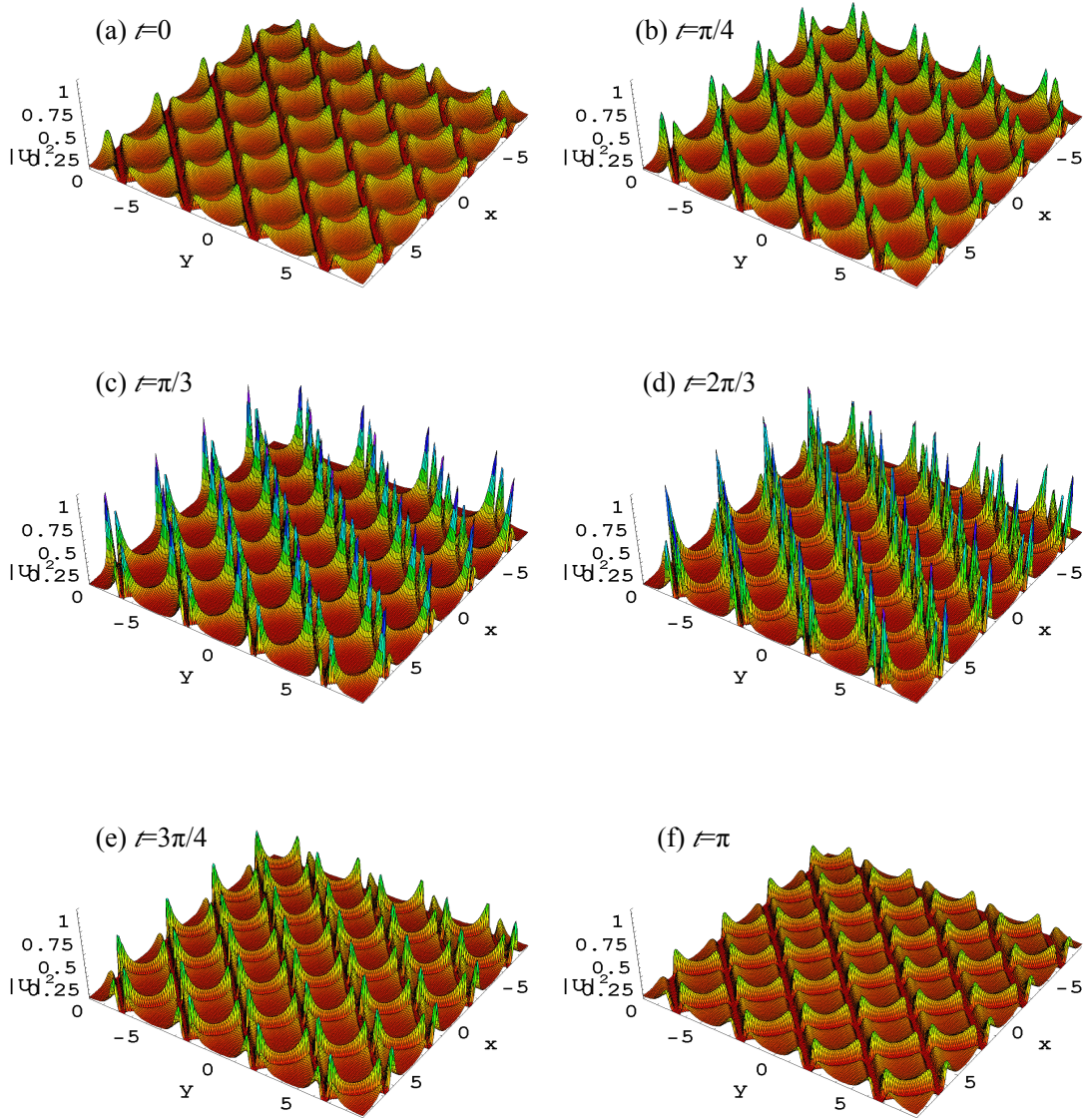

Fig. 4. (Color online) Periodic wave pattern. Their parameters are $a_0 = 2, a_1 = 1, a_2 = 2, a_3 = 2$.

**Case 4:** Double-instantons excitation

An instanton is a classical solution to the equations of motion of the classical field theory on a Euclidean space-time. Instantons are important in quantum field theory, they appear in the path integral as the leading quantum corrections to the classical behavior of a system, furthermore, they are topologically nontrivial solutions of Yang-Mills equations that absolutely minimize the energy functional within their topological type, they can be used to study the tunneling behavior in various systems such as a Yang-Mills theory [27].

The Fig. 5 shows a double-instanton of DS system, which can be obtained by

$$\begin{cases} p(\zeta,t) = \exp\{\zeta + 2t + 1\} + \exp\{\zeta + t + 1\} + \exp\{-(\zeta^3 + 1)^{-1} + 2t + 1\}, \\ q(\eta,t) = \exp\{\eta + 2t + 1\} + \exp\{\eta + t + 1\}. \end{cases} \quad (18)$$

From Fig. 5 (a) (b) (c), both three-dimensional pictures and their projections, we can see the amplitudes of the field $U$ diminish with the time $t=0, t=3$ and $t=6$. Furthermore, from the projections, we can notice that two instantons are bonded with together, that is we called a double-instanton.

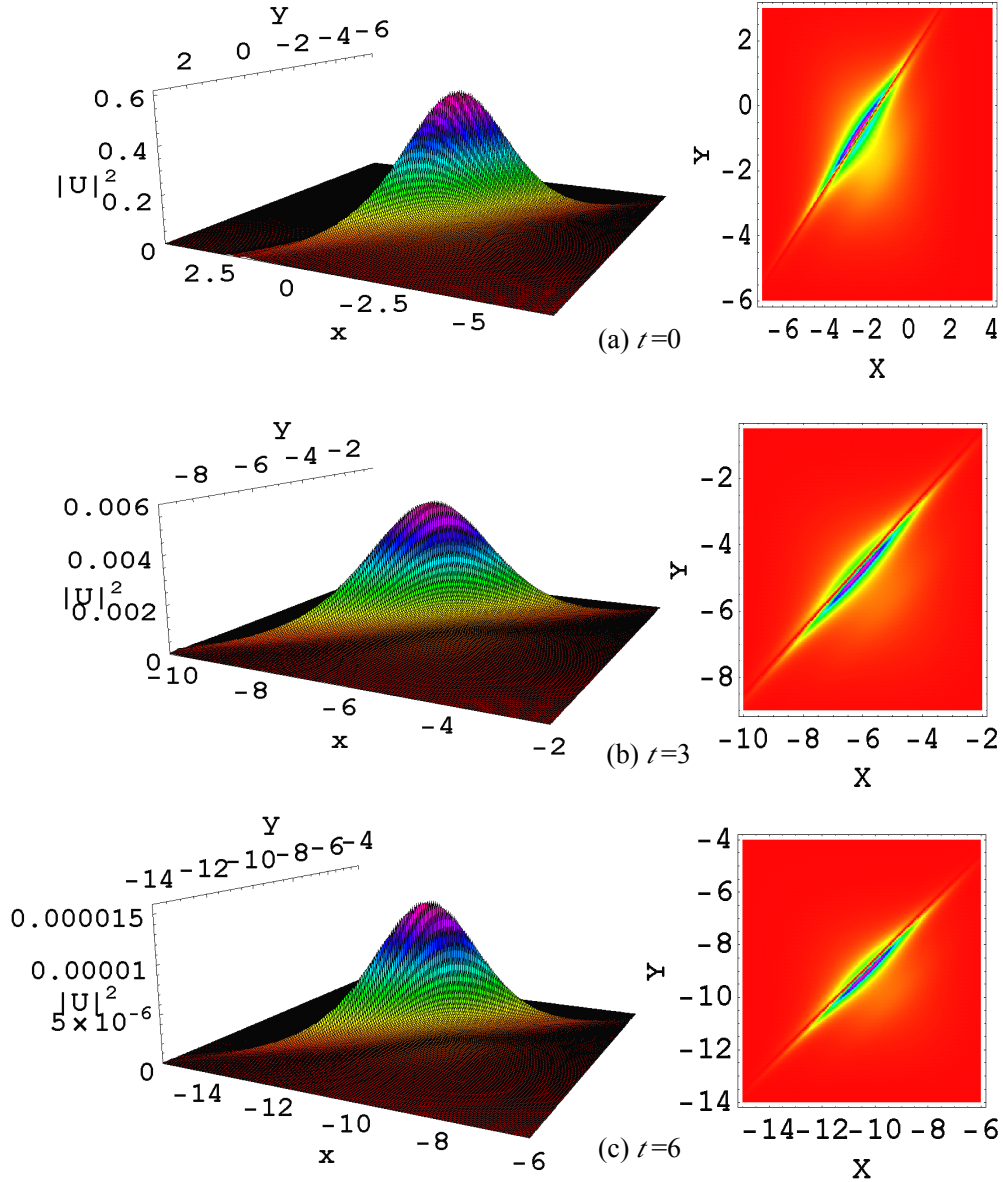

Fig. 5. (Color online) Double-instanton excitations. Their parameters are $a_0 = 2, a_1 = 1, a_2 = 2, a_3 = 2$.

Those ranges of field $U$ are $\{0, 0.6\}, \{0, 0.006\}$ and $\{0, 0.000015\}$, respectively.

# IV. Conclusions and Discussions

In this paper, the variable separation approach is used for searching solutions of Davey-Stewartson (DS) equations which contain high nonlinear and gain terms. It can be concluded that the variable separation approach is a very powerful and efficient method in finding exact and explicit solutions for wide classes of problems. In particular, we have given several excitations such as dromions, solitoffs, breathers, periodic wave patterns and double-instantons for the DS equations. In addition, we also have discussed several characters of these different excitations.

Our analytical study of solutions can explain many experimental results. Due to the DS equations by now arising in many different kinds of situations, these different excitations of solution may have a wide range of applicability. In the past few decades, solitons appear widely in solid, fluid and optical media. Nonlinear localized structures have received much attention and taken an important role in the study of Bose-Einstein condensation (BEC) [28] and matter waves.

Refs. [29, 30] have demonstrated that Faraday pattern waves and nonlinear interactions are created in BEC, and showed an initial transverse breathing mode excitation of the condensation leading to spontaneous pattern formation in the longitudinal direction. Analysis of the breather in this paper clearly illustrates this similar excitation (a standing breather corresponds to localized solutions whose amplitudes vary with time) and its characteristics. The exact numerical instanton solutions [31], which describe quantum tunneling from both the metastable and radially excited states of BEC, demonstrate the value of the instanton excitations in the analysis of BEC. In the study of the nonlinear optical systems, Ref. [32] exhibited the experimental control of drifting solitary structures and their motions by using an incoherent external amplitude control in a unique way. And the drifting solitons can be explained in this paper by the shape changes of the soliton solutions with time, such as dromion excitation (Fig. 2), solitoff excitation (Fig. 3), periodic wave pattern excitation (Fig. 4), etc. And the excitation patterns discussed above can also be applied to the plasmas, shallow sea area waves and the other branches of physics.

In summary, we have studied the general (2+1)-dimensional DS equations with nonlinear and gain terms, we found some important excitations through variable separation approach. In particular, we have analyzed these main excitations, which would be important to the development of modern physics. We predict that similar solutions and excitations will also exist in other related revolution equations such as nonlinear Schrödinger equation, and more analysis and details will be left for our future studies.

**ACKNOWLEDGMENT**

The work is supported by the Project of Knowledge Innovation Program (PKIP) of Chinese Academy of Sciences ( Grant No. KJCX2.YW.W10 ), National Natural Science Foundation of China (Grant No. 10875009 ) and Beijing Natural Science Foundation (Grant No. 1072005).

**References**
[1]  M. Tajiri, and H. Maesono, Phys. Rev. E **55**, 3351 (1997)
[2]  S. Ghosh, and S. Nandy, Nucl. Phys. B **561**, 451 (1999)
[3]  S. Nandy, Nucl. Phys. B **679**, 647 (2004)

[4] S. Deng, and Z. Qin, Phys. Lett. A **357**, 467 (2006)

[5] J. Liu, and Y. Li, Comput. Phys. Commun. **179**, 724 (2008)

[6] A. Wazwaz, Appl. Math. Comput. **200**, 160 (2008)

[7] R. Hirota, (Cambridge University Press, Cambridge, 2004)

[8] G. Xu, Chaos, Solitons & Fractals **30**, 71 (2006)

[9] B. Tian, and Y. Gao, Appl. Math. Comput. **84**, 125 (1997)

[10] A. Davey, and K. Stewartson, Proc. Roy. Soc. London Ser. A **338**, 101 (1974)

[11] M. Tajiri, and T. Arai, Phys. Rev. E **60**, 2297 (1999)

[12] M. Tajiri, K. Takeuchi, and T. Arai, Phys. Rev. E **64**, 56622 (2001)

[13] M. Tajiri, H. Miura, and T. Arai, Phys. Rev. E **66**, 67601 (2002)

[14] K. W. Chow, and S. Y. Lou, Chaos, Solitons & Fractals **27**, 561 (2006)

[15] X. Y. Tang, K. W. Chow, and C. Rogers, Chaos, Solitons & Fractals **42**, 2707 (2009)

[16] A. Eden, S. Erbay, and I. Hacinliyan, Chaos, Solitons & Fractals **41**, 688 (2009)

[17] X. Y. Tang, S. Y. Lou, and Y. Zhang, Phys. Rev. E **66**, 46601 (2002)

[18] R. Hirota, (Cambridge University Press, Cambridge, 2004)

[19] Y. C. Huang, X. G. Lee and M. X. Shao. Mod. Phys. Lett. A**21,** 1107 (2006); Y. C. Huang and Q. H. Huo, Phys. Lett. B **662**, 290 (2008)

[20] Y. C. Huang, C. X. Yu, Phys. Rev. D **75**, 044011 (2007); L. Liao and Y. C. Huang, Ann. Phys. (N. Y.) **322**, 2469 (2007)

[21] Y. C. Huang and G. Weng, Commun. Theor. Phys. **44,** 757(2005); Y. C. Huang, L. Liao and X. G. Lee, European Physical Journal C **60,** 481 (2009)

[22] Y. C. Huang and B. L. Lin, Phys. Lett., A**299**, 644(2002)

[23] P. G. Kevrekidis, A. Saxena, and A. R. Bishop, Phys. Rev. E **64**, 26613 (2001)

[24] K. Nozaki, Phys. Rev. Lett. **49**, 1883 (1982)

[25] J. Dorignac *et al.*, Phys. Rev. Lett. **93**, 25504 (2004)

[26] A. Trombettoni, and A. Smerzi, Phys. Rev. Lett. **86**, 2353 (2001)

[27] M. Eto *et al.*, Phys. Rev. Lett. **95**, 252003 (2005)

[28] Y. Kagan, and L. A. Maksimov, Phys. Rev. Lett. **85**, 3075 (2000)

[29] P. Engels, C. Atherton, and M. A. Hoefer, Phys. Rev. Lett. **98**, 95301 (2007)

[30] K. Staliunas, S. Longhi, and G. A. J. de Valcarcel, Phys. Rev. A **70**, 11601 (2004)

[31] J. Skalski, Phys. Rev. A **65**, 33626 (2002)

[32] C. Cleff, B. O. Gutlich, and C. Denz, Phys. Rev. Lett. **100**, 233902 (2008)